\documentclass[usenatbib,usegraphicx]{mn2e}
\usepackage{amsmath,amssymb,lscape,nth,relsize,algorithm2e,tikz,soul,bm}
\title[Symplectic integration for the $N$-body problem]
{Symplectic integration for the collisional gravitational $N$-body problem}

\author[David M. Hernandez and Edmund Bertschinger]
	{David M. Hernandez \thanks{Email: dmhernan@mit.edu (DMH); edbert@mit.edu (EB)} and Edmund Bertschinger \footnotemark[1]  \\
	Department of Physics and Kavli Institute for Astrophysics and Space Research,
	Massachusetts Institute of Technology, 77 Massachusetts Ave., \\ Cambridge, Massachusetts 02139, USA\\
	}
	\date{Accepted 2015 June 26.  Received 2015 June 3; in original form 2015 March 13}
\begin{document}

\maketitle

\label{first page}
\begin{abstract}
We present a new symplectic integrator designed for collisional gravitational $N$-body problems which makes use of Kepler solvers.  The integrator is also reversible and conserves 9 integrals of motion of the $N$-body problem to machine precision.  The integrator is second order, but the order can easily be increased by the method of \citeauthor{yos90}.  We use fixed time step in all tests studied in this paper to ensure preservation of symplecticity.  We study small $N$ collisional problems and perform comparisons with typically used integrators.  In particular, we find comparable or better performance when compared to the 4th order Hermite method and much better performance than adaptive time step symplectic integrators introduced previously.  {We find better performance compared to SAKURA, a non-symplectic, non-time-reversible integrator based on a different two-body decomposition of the $N$-body problem.}  The integrator is a promising tool in collisional gravitational dynamics. { }
\end{abstract}


\section{Introduction}
Astrophysical $N$body problems can be classified by their collisional nature.  We define collisionless problems as those where the two body relaxation time (see \cite{bin08}) is larger than the timescale of interest.  A typical globular cluster has $t_{\mathrm{relax}} \approx 400$ Myrs, whereas a dark matter (DM) halo's relaxation time is many orders of magnitude larger than a Hubble time.  If we are concerned with evolution over the time scale of the Universe, the former is collisional while the later is collisionless.  In the collisionless case, a test particle's motion is approximately due to a smooth potential of the mass distribution.  In this case the dynamics is given by the collisionless Boltzmann equation, a partial differential equation in $6$ variables.  Collisional problems are much more complicated, requiring a solution of the $6N$ dimensional Liouville equation, with $N$ the particle number.  

DM is typically believed to be collisionless.  Collisionless problems, like those in DM simulations, are typically solved by a Monte Carlo sampling of $N$ particles of the phase fluid.  To avoid non-physical two-body {encounters and two-body relaxation}, researchers usually include a softening length such that the force goes to a finite constant as inter particle distances go to 0 (see for example \cite{vog08}).  The dynamics then remain governed approximately by a collisionless Boltzmann equation.  However, too high a softening length leads to other errors \citep{bert98}.  {A study of errors in dynamics introduced by softening lengths is carried out in} \cite{deh01}.   The impact of the softening length cannot be overstated: without this feature, modern codes would not be able to carry out the  large DM $N$-body integration with $N > 10^9$.

We cannot include a softening length in collisional problems, such as in globular clusters, where close particle encounters drive the dynamics.  Tight binaries, hierarchical triples, and higher order systems will form.  In order to model these systems accurately, standard integrators will require computationally prohibitive resources in the form of small time steps \citep{kv00}.  These new systems also introduce new dynamic timescales, and standard integrators will not be suited for the resulting large dynamic range of timescales of the problem.

The most popular work around these problems is regularization, described in \cite{aar08}.  The basic idea is to perform a coordinate and time transformation such that in the new coordinates, the equations of motion remain nonsingular as the inter-particle separation goes to 0.  There are ways to regularize three (and more) body encounters.  

With or without regularization, most integrators up until now show inaccuracies when applied for long times to general collisional systems.  One way of mitigating this problem is to incorporate a \textit{geometrical} integrator \citep{hair06, leim04}, an integrator that conserves qualitative features of the original problem.  Such features can be due, for example, to continuous symmetries in the Hamiltonian, leading to integrals of motion, or discrete symmetries such as time reversibility.  Symplectic integrators are one such geometric integrator which preserve phase space volume.  The advantages of such an integrator can be powerful for long term evolution of problems due to a lack of secular drift in energy errors and preservation of global structure in the governing differential equations.  For this reason they are the preferred method in cosmological simulations \citep{spr05} and planetary simulations \citep{wis91}, but have not been successfully employed yet in globular clusters or other collisional systems.  

Symplectic integrators have not found success in their application to collisional systems in the past because when time steps are adapted as a function of phase space or if the integrator changes form as a function of phase space (described as a ``switch'' in \cite{leim04}), the integrators no longer approximately obey a Hamiltonian, a major feature which helps make them successful in the first place.  Nonetheless, \cite{mik99} and \cite{preto99} independently discovered a symplectic integrator for small $N$ collisional problems which is able to adapt its time step as a function of the total potential energy of the system.

 Time reversible integrators share similarities to some symplectic integrators for regular or near regular motion \citep{hair06} and indeed time reversible methods for collisional systems have been developed.  Especially noteworthy is an adaptive symmetric Hermite method \cite{kok98} implemented in the popular Aarseth codes \citep{aar08}, but the computational cost of the integrator is high and its use is recommended only for planetary systems.  The leapfrog method has also been made to be adaptive and time reversible: this integrator is studied in \cite{gon14} and other works, but its performance is not found to be as favorable as other existing methods.

In this paper, we present a new symplectic integrator for collisional gravitational $N$-body dynamics.  The integrator is inspired by the non-symplectic and non-reversible integrator in \cite{gon14}, {SAKURA}, and makes use of Kepler solvers.  {Like SAKURA we decompose the $N$-body problem into two-body problems.  In contrast to SAKURA, our two-body problems are not independent}.  The integrator is reversible and symplectic and conserves 9 integrals of motion of the $N$-body problem to machine precision.  The integrator is second order, in the sense of \cite{hair06}, but the order can be increased by the method of \cite{yos90}.  We use fixed time step in all tests studied in this paper in order to remain exactly symplectic.  We study small $N$ collisional problems and perform comparisons with typically used integrators.  In particular, we find comparable or better performance when compared to the 4th order Hermite method and much better performance than the second order symplectic integrator introduced by \cite{mik99} and \cite{preto99}.  {We also find improved performance over SAKURA. Thus our} integrator is promising as a tool in collisional gravitational dynamics.  We plan larger $N$ tests of the method in future work.

The organization of the paper is as follows.  In Section \ref{sec:bask} we outline the method along with its general properties and give first illustrations of its power in handling collisional problems.  Section \ref{sec:deriv} provides a mathematical derivation of the method along with a proof of its symplecticity and other properties.  Section \ref{sec:symptild} shows numerical evidence for symplecticity and the existence of a Hamiltonian closely followed by the method.  

Section \ref{sec:numtest} is dedicated to numerical tests.  Section \ref{sec:reg} studies tests of regular 3 body systems in order to clearly show the advantages of the symplecticity of the method over other commonly used integrators.  In Section \ref{sec:chaos} we study the more common chaotic $N$-body problem and show the good performance of the method persists over other integrators.  {We also argue why symplectic methods are still highly desirable for chaotic problems.}  Finally, Section \ref{sec:comb} shows how to combine the method with a traditional leapfrog method while maintaining symplecticity and reversibility.    

\section{The method}
\subsection{Basic use and overview}
\label{sec:bask}

First we outline the steps for the basic integrator in pseudocode, a map of phase space coordinates $\phi_h^2$.  It is a symplectic, symmetric {( or reversible)} second order integrator, although it is easy to generalize to other orders via the method of \cite{yos90}.  First we begin with a first order mapping, $\phi_h$.  A drift is defined for a coordinate system in which the $N$-body problem is separable: the kinetic energy depends only on momentum and the potential energy only on coordinates.  For example, a drift in cartesian coordinates is $\bm{x}^\prime = \bm{x} + h \bm{v}$, where $\bm{x}^\prime$ is a new position, and $\bm{x}$ and $\bm{v}$ are the old position and velocity respectively, and $h$ is a time step.

\begin{algorithm} 
\caption{A Kepler integrator mapping $\phi_h$.}
\label{alg:phih}
     \SetAlgoLined
     Drift all particles for time $h$ \;
     \For{ all pairs of particles ($i$, $j$)}{
 		Drift particles $i$ and $j$ for time $-h$ \;
		Apply a Kepler solver to advance the relative coordinates of $i$ and $j$ by $h$ \;
		Advance center of mass coordinates of $i$ and $j$ by $h$ \;
		}
\end{algorithm}
Now we define an integrator that has the same steps but exactly in reverse as $\phi_h^\dag$.  Our second order integrator is then their composition:
\begin{equation}
\phi_h^2 = \phi_{h/2}^\dag \phi_{h/2}
\label{eq:phi2}
\end{equation}
There is no preferred order of particles in the loop of Algorithm \ref{alg:phih} in general.  The order probably affects the specific form of higher order terms in the Hamiltonian obeyed by the method (described in Section \ref{sec:deriv}) we did not calculate, but we did not find significant differences in accuracy when changing the order in which particles are updated for different problems.  There could be special problems for which a specific ordering gives better results.
The composition of the two methods makes the map time reversible, as will be discussed below.  With the two compositions there are $N(N-1)$ total Kepler solvers, where $N$ is the number of particles.  These Kepler solvers account for the only significant computation time.
It is possible to create a second order integrator with half as many Kepler solvers (in fact, Algorithm \ref{alg:phih} is such an example), but this is the simplest we could find that is symmetric, and thus time reversible.  It can be shown theoretically that a reversible integrator alone already shares many properties of a symplectic integrator when applied to integrable and near-integrable systems (Chapter XI \cite{hair06}).  While the $N$-body problem is not integrable or near-integrable, we can expect advantages in structure preservation by using a reversible method.  

We can show simply $\phi_h^2$ is reversible by checking definition V.1.4 in \cite{hair06}.  For a reversible map $\phi_h$, 
\begin{equation}
\phi_h \phi_{-h} = \bm{I},
\label{eq:revers}
\end{equation}
where {$\bm{I}$} is the identity matrix.  A numerical method should obey eq. \ref{eq:revers}, independent of the step size $h$, to round off error.{\footnote{Actually, for several reversible integrators we tested, the error grows for $h \ge 1$.  This is due to errors in the finite precision arithmetic and a loss of decimals stored.}}  We find $\phi_h^2$ has such a property.  We note here that, contrary to their claim, we do not find the method of \cite{gon14}, which also provides an integrator with Kepler solvers, to be time reversible.  Tests show that the error in eq. \ref{eq:revers} applied to their method grows with $h$ and is not at the level of roundoff.  {We prove SAKURA's irreversibility and explain why SAKURA is not reversible in Appendix} \ref{app}.

When compared to a popular Drift-Kick-Drift (DKD) leapfrog method (see the introduction of \cite{preto99} for a description of DKD and its counterpart KDK), as far as the main computational work is concerned, we have substituted force calculations between the ${N(N-1)}/2$ pairs for solving ${N(N-1)}/2$ Kepler problems.  Generally, the Kepler solvers will take slightly longer than the force calculations.  We can compare the computational cost of solving a two body problem (2BP) between $\phi_h^2$ and DKD.  For the 2BP, $\phi_h^2$ reduces to a two body Kepler solver, assuming the center of mass is stationary.  {A two body Kepler solver advances 6-dimensional relative phase space coordinates.}    

To test this, we choose reduced mass $\mu=1$, total mass $M=1$, eccentricity $e=0.9$, semi-major axis $a=1$ (so that the period is $2\pi$), and time step $h=0.01$ in units where $G=1$.  The initial conditions are at apoapsis: $x=1+e$, $y=0$, $v_x = 0$, $v_y = \sqrt{(1-e)/(1+e)}$.  Running the 2BP for 25 orbits on a Macintosh laptop computer, the computation time for this problem is about $38\%$ longer for the Kepler solver (for equal $h$) {compared with DKD}.  Our Kepler solver uses universal variables for advancing the general Kepler problem with bound or unbound orbits following \cite{dan88}; see also \cite{mik99c}.  

Map $\phi_h^2$ is symplectic as we will prove analytically and show numerically.  We will also show numerically that of the $10$ integrals of motion known for the general $N>3$ $N$-body problem, $\phi_h^2$ conserves 9 to machine precision.  As for the last integral of motion, the energy, a surrogate Hamiltonian $\tilde{H}$ is conserved that differs from the $N$-body Hamiltonian by the order $h^2$, because it is a second order integrator.  It will remain close to the energy as long as $h$ is small enough, especially in a near regular system.  For rigorous results on this statement, see \cite{hair06} and \cite{engle05}.  The DKD leapfrog integrator, in addition to being symplectic and reversible, also has the same conservation properties.  However, it is not as well suited for collisional studies as $\phi_h^2$; the explanation for this lies in the behavior of the infinite series in $h$ found in its $\tilde{H}$: specifically, theorems for long term energy conservation presented in \cite{hair06} no longer apply.

We illustrate this point in Table \ref{tab:dkdvssak}.  Here we test the a three body problem with the two integrators, DKD leapfrog and $\phi_h^2$.  We choose the three body Pythagorean problem (3BPP), described in detail in \cite{sze67}, with their same initial conditions.  We use time step $h=0.01$.  For reference, by time $t=10$, there have been 6 close encounters.  The first close encounter happens shortly after $t=1.5$.     

We run the integrators for $t = 2.0$, a time shortly after the first close encounter.  Even when $\phi_h^2$ is given a time step 15 times larger, and its computation time is more than 5 times smaller, it performs similarly to a DKD integrator.  This test shows the unsuitability of DKD for collisional studies.  We see conservation of all integrals of motion of this problem: the energy, angular momentum, and four center of mass coordinates.  Section \ref{sec:deriv} explains the details of why $\phi_h^2$ has these conservation properties.  We will revisit in Section \ref{sec:chaos} the unsuitability of leapfrog for use in chaotic problems.

\begin{center}
\begin{table*}
\caption{Comparison of performance of two second order methods: DKD leapfrog integrator vs $\phi_h^2$ for a collisional study of the 3BPP described in the text.  $t_{max}$ is $2.0$, shortly after the first close encounter.  Even when $\phi_h^2$ is given a time step nearly 10 times larger, and its computation time is more than 5 times smaller, it performs at least as well as a DKD integrator.}
\centering
\begin{tabular}{| c || c| c| c| c| c| c| c| c |}
	\hline
	 & $\left|\Delta E/E_0 \right|$ & $dL$ & $dx_{cm,1}$ & $dx_{cm,2}$ & $dp_{cm,1}$ & 
	$dp_{cm,2}$ &  $t_{\mathrm{cpu}}$  & $h$\\ [3ex] \hline
	DKD leapfrog & $8.2 \times 10^{-6} $& $1.2\times 10^{-13}$ & $8.4 \times 10^{-15}$ & $1.3 \times 10^{-13}$& $2.0 \times 10^{-13}$ & $2.1 \times 10^{-14}$ & $7.5$ & $0.0001$\\ \hline
  $\phi_h^2$ & $3.7 \times 10^{-6} $& $1.1\times 10^{-13}$ & $4.7 \times 10^{-14}$ & $1.4 \times 10^{-14}$& $2.6 \times 10^{-14}$ & $8.0 \times 10^{-15}$ & $1.4$ & $0.0015$ \\ \hline
	\end{tabular}

\label{tab:dkdvssak}
\end{table*}
\end{center}   

\subsection{Derivation of method and basic properties}
\label{sec:deriv}
This Section and Section \ref{sec:symptild} derive and prove the conservation properties of $\phi_h^2$.  Readers not interested in the technical details may wish to skip to the performance of the integrator when compared to other integrators in Section \ref{sec:numtest}.

To derive $\phi_h^2$, first proceed as in Section \ref{sec:bask}.  We first build {a} second order integrator $\phi_h$ from Algorithm \ref{alg:phih}.  We will follow notation that is conventional in \cite{yos90}, \cite{hair06}, and others.  We motivate the integrator by writing first the $N$-body Hamiltonian,
\begin{equation}
\begin{aligned}
H &= T + V \\
&= T + \sum_i \sum_{j > i} V_{i j} \\
&= T + \sum_i \sum_{j > i} \left( K_{i j} - T_{i j} \right).
\end{aligned}
\label{eq:Hamilt}
\end{equation}
Here $T$ is the kinetic energy, $V$  is the potential energy from Newton's gravitational law, $V_{i j}$ is the pairwise potential of particles, $T_{i j}$ is the kinetic energy of particle $i$ plus particle $j$, and $K_{i j}$ is the two body Hamiltonian for particles $i$ and $j$ ($K$ stands for Kepler) {including both relative and center of mass terms}.  

Using an operator splitting method (see \cite{yos90}), we construct
\begin{equation}
\begin{aligned}
\phi_h &=  \left( \prod_{(i,j)~ pairs} \exp \left({h D_{K_{i j}}} \right) \exp \left({- h D_{T_{i j}}} \right) \right)  \exp \left({h D_T} \right). \\
&=  \left( \prod_{(i,j) ~pairs} \psi_h^{3ij}  \psi_{-h}^{2ij} \right) \psi_h^1,
\end{aligned}
\label{eq:firsto}
\end{equation}
where each $\psi$ mapping is defined as one of the exponentials.  Here $D_{A}$ is defined by $D_{A} f = \{f,A\}$, where $\{\}$ are Poisson brackets.  Exponentials are calculated by writing their Taylor series.  Thus we can check that mapping $\psi_h^1$ corresponds to a drift of time $h$, defined in Section \ref{sec:bask}.  {Equation} \ref{eq:firsto} {corresponds exactly to Algorithm} \ref{alg:phih}.  As mentioned before, we found no generally preferred ordering in the product of eq. \eqref{eq:firsto}.  Note that the $\psi$ in parenthesis are invariant under exchange of $i$ and $j$.  

First we prove the symplecticity of map \eqref{eq:firsto}.  Because a composition of symplectic mappings is symplectic, we will show each exponential mapping in eq. \eqref{eq:firsto} is symplectic.  In {the} canonical coordinate basis {discussed below}, the symplectic matrix is defined by 
\begin{equation*}
\bm{\Omega} = 
	\begin{bmatrix}
	{0} & {\bm{I}} \\
	-{\bm{I}} & {0} \\
	\end{bmatrix}.
\end{equation*} 
$\bm{I}$ and ${0}$ are the $3N \times 3N$ identity and zero maps respectively.  For a map with associated Jacobian ${\bm{J}}$ and Jacobian transpose ${\bm{J}}^\intercal$, the symplectic condition says \citep{suss01}
\begin{equation}
\bm{\Omega} = {\bm{J}}^\intercal {\bm{\Omega}} {\bm{J}}. 
\label{eq:sympcond}
\end{equation}
We check this condition for eq. \eqref{eq:firsto}, starting with $\psi_h^1$.  We can calculate the Jacobian resulting from this map acting on a phase space $\bm{y}$, where $\bm{y}$ is a $6N$ vector composed of the coordinates and momenta, in that order.  We use cartesian coordinates and their associated linear momenta from the $N$-body Lagrangian.    
${\bm{J}}$ is given by
\begin{equation}
 {\bm{J}} = 
 \begin{bmatrix}
	\bm{I} & \bm{M}\\
	0 & \bm{I}\\
\end{bmatrix}.
\label{eq:jacobian}
\end{equation}
Here, $\bm{M}$ has the property 
\begin{equation}
\bm{M}^{\intercal} = \bm{M}.
\label{eq:mtranspose}
\end{equation}
We can now verify the symplectic condition \eqref{eq:sympcond}:
\begin{equation*}
\begin{aligned}
{\bm{J}}^\intercal {\bm{\Omega}} {\bm{J}} &= 
\begin{bmatrix}
	0 & \bm{I}\\
	-\bm{I} & \bm{M}^\intercal-\bm{M}\\
\end{bmatrix}
\\
&= {\bm{\Omega}}.
\end{aligned}
\end{equation*}
This proves the familiar result that a drift of all particles is a symplectic map.

Next we focus on $\psi_h^{2ij}$, a drift of particles $i$ and $j$.  The associated Jacobian again has form eq. \eqref{eq:jacobian} with property \eqref{eq:mtranspose} so map $\psi_h^{2ij}$, a drift of particles $i$ and $j$ is symplectic.

Finally, we have the map $\psi_h^{3ij}$.  By definition, this mapping is a Hamiltonian flow of particles $i$ and $j$ and thus symplectic with respect to their 12 dimensional phase space.  It is easy to show then using condition \eqref{eq:sympcond} that the Jacobian of the entire phase space vector $\bm{y}$ is also symplectic.  Since we have shown $\psi_h^1$, $\psi_h^{2ij}$, and $\psi_h^{3ij}$ are symplectic maps, we have proven $\phi_h$ is a symplectic map.  

We discuss the integrals of motion associated with the continuous symmetries of the Hamiltonian obeyed by $\phi_h$.  We are first able to show that the total angular momentum is conserved exactly to machine precision.  To do this, we can check that $\psi_h^1$, $\psi_h^{2ij}$, and $\psi_h^{3ij}$ conserve the total angular momentum.  The 6 center of mass integrals are also conserved exactly by these maps.  It is easy to check the conservation properties analytically, and numerical experiments support the finding.  

DKD leapfrog shares the property with $\phi_h$ that 9 of the 10 integrals of motion are conserved exactly{, and it is symplectic.  DKD is a well known integrator, but let us discuss its properties carefully as they will prove useful for Section} \ref{sec:comb} {.  We have already studied the drift maps D and all that remains for analysis is the kick map K.}

{We can represent K as $\exp(h D_V)$.  $D_V$ is a sum of pair-wise potential energy operators that commute because the potential energies depend only on position.  Thus, $\exp(h D_V)$ can be written as a product of maps of form $\exp(h D_{V_{i j}})$.  Let us study the map $\exp(h D_{V_{i j}})$.  This map shifts the momentum of particles $i$ and $j$ by a term proportional to their mutual force.  The Jacobian of K has a form similar to those we have seen:}
\begin{equation}
 {\bm{J}} = 
 \begin{bmatrix}
	\bm{I} & 0\\
	\bm{M} & \bm{I}\\
\end{bmatrix},
\label{eq:jacobian2}
\end{equation} 
{with $\bm{M}$ a symmetric matrix.  Then, by similar arguments to those following equation} \eqref{eq:jacobian} {, K is symplectic.  The conservation of total momentum follows from Newton's third law.  Because position coordinates remain unaffected by K, conservation of the remaining center of mass constants follows.  The total angular momentum is conserved because the change in momenta is a function only of position. }
      
We see that a discrete symmetry of the problem, time reversibility, is not respected by $\phi_h$.  It is obvious from inspection of $\phi_h$ that eq. \eqref{eq:revers} does not hold and one can indeed verify this numerically by running the integrator forwards and backwards for a time step.  We should try to correct this feature since, as noted previously, time reversibility under some circumstances provides benefits similar to those of symplecticity.

When one tests the order of $\phi_h$, one finds it to be a second order integrator.  Where does this come from?  It comes from a simplification in the final integral of motion of $\phi_h$, a surrogate Hamiltonian $\tilde{H}$.  We will investigate where $\tilde{H}$ comes from now.  To begin, we state that the exact Hamiltonian canonical coordinate transformation after time $t$ is given by  
\begin{equation}
\bm{y}(t) = \exp \left(t D_H \right) \bm{y}(0).
\label{eq:hamiltevol}
\end{equation}
Eq. \eqref{eq:hamiltevol} follows from the Hamiltonian equations of motion {for a time-independent Hamiltonian}.  Now, we write $\phi_h = \exp \left(h D_{\tilde{H}} \right)$ (note we are using a $\tilde{H}$ instead of $H$ operator).    Then we use eq. \eqref{eq:firsto} with the Baker-Campbell-Hausdorff formula (see \cite{hair06}) to obtain a relationship between $D_{\tilde{H}}$, $D_{K_{i j}}$, $D_{T_{i j}}$, and $D_T$.  Next, we use the following relation, which can be verified directly:
\begin{equation*}
[D_F,D_G] = D_{\{G,F\}},
\end{equation*}
with $[]$ referring to commutation.  With this relationship, we can solve for $\tilde{H}$: 
\begin{equation}
\begin{aligned}
\tilde{H} &= H + h/2 \left( - \left\{V,T \right\} + \sum_i \sum_{j > i} \left\{V_{i j}, T_{i j} \right\} \right)+\mathcal O(h^2) \\
&= H + \mathcal O(h^2),
\label{eq:pertham}
\end{aligned}
\end{equation}
i.e. the order $h$ term vanishes exactly.  This unexpected result explains why the order of $\phi_h$ is actually 2 numerically.  Illuminating theoretical and numerical work on $\tilde{H}$ can be found in \cite{engle05} and \cite{hair06}.  Even though the infinite series in $\tilde{H}$ generally diverges, {they show} the error in conservation of the first $N${ terms in $\tilde{H}$, with $N$ determined by $h$,} is exponentially {suppressed} over exponentially long times.  Because the 0th order of $\tilde{H}$ is the energy, the long term conservation of energy follows.  

$\phi_h$ is second order, and conserves 9 out of 10 integrals of motion exactly.  But we can do better and require time reversibility, albeit at the cost of twice the number of Kepler solvers (two Kepler solvers per pair of particles).  To do this we first introduce the adjoint map as defined in \cite{hair06}: 
\begin{equation}
\zeta_h^\dag = \zeta_{-h}^{-1}.
\label{eq:adjoint}
\end{equation}
\cite{hair06} {shows that} adjoint methods can be used to increase the order of an integrator, but this will not work for $\phi_h$.  Adjoint methods have other uses: for a reversible method we have 
\begin{equation*}
\zeta_h = \zeta_h^\dag,
\end{equation*}
an equation we can see, by inspection, does not hold for $\phi_h$.  But the map $\phi_h^2 = \phi_{h/2}^\dag \phi_{h/2}$ will be symmetric (while still incorporating the other conservation properties discussed above).  $\phi_h^2$ has another property not shared by $\phi_h$ in the form of $\tilde{H}$.  Because $\phi_h^2$ is symmetric, its $\tilde{H}$ will only have even terms in $h$ (for a proof, see \cite{yos90}), specifically:
\begin{equation}
\tilde{H} = H + f(\bm{y}) h^2 + g(\bm{y}) h^4 + \ldots,
\label{eq:perth}
\end{equation}    
where $f$ and $g$ are known functions of phase space.  This is why symplectic and reversible integrators can only be even order.

\subsubsection{Symplecticity and the perturbed Hamiltonian $\tilde{H}$}
\label{sec:symptild}
We first wish to study the properties of $\tilde{H}$.  It is hard to analytically calculate $f$ and $g$ from above but an easier alternative exists.  Consider an alternate symplectic, symmetric, second order map, slightly more complicated than $\phi_h^2$,
\begin{equation*}
\zeta_h^2 = \zeta_{h/2}^\dag \zeta_{h/2}
\end{equation*}
with
\begin{equation}
\zeta_h = \left( \prod_{(i,j)~ pairs} \psi_{-h/2}^{2ij} \psi_{h}^{3ij}  \psi_{-h/2}^{2ij} \right) \psi_{h}^1.
\label{eq:zetah}
\end{equation}
The difference with $\phi_h$ is that there are twice as many drift steps.  We write the pseudocode for integrator eq. \ref{eq:zetah} for clarity in Algorithm \ref{alg:zeta}.  
\begin{algorithm} 
\caption{Pseudocode for $\zeta_h$ defined by eq. \ref{eq:zetah}.}
\label{alg:zeta}
     \SetAlgoLined
     Drift all particles for time $h$ \;
     \For{ all pairs of particles ($i$, $j$)}{
 		Drift particles $i$ and $j$ for time $-h/2$ \;
		Apply a Kepler solver to advance the relative coordinates of $i$ and $j$ by $h$ \;
		Advance center of mass coordinates of $i$ and $j$ by $h$ \;
		Drift particles $i$ and $j$ for time $-h/2$ \;
		}
\end{algorithm}
{Using a Kepler map sandwiched between two drift maps in equation} \eqref{eq:zetah} {reminds us in form to an integrator presented in} \cite{rein11}, {equation 17.}  \cite{rein11} {studied the three-body problem in Hill's approximation and solved for the relative coordinates of the small masses in a rotating frame.}

\begin{figure}
\includegraphics[width=.5\textwidth]{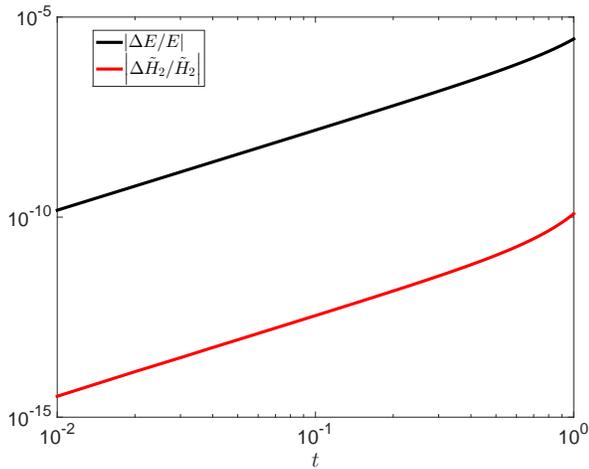}
  \caption{Plot showing the improved conservation of $\tilde{H}_2$ as compared to conservation in energy $E$, as expected for small enough $h$.  We use the map $\zeta_h^2$ and its associated $\tilde{H}_2$, applied to the 3BPP run for time $t=1.5$ with $h=.01$, before any collisions occur.}
  \label{fig:tildeh}
  \centering
\end{figure}

The symmetric nature of the terms within parenthesis makes computing $\tilde{H}$ easier for $\zeta_h^2$ {than for $\phi_h^2$}.  The result is as follows:
\begin{equation}
\tilde{H} = \tilde{H}_2 + \mathcal O(h^4)
\label{tildhzeta}
\end{equation}
with
\begin{equation*}
\begin{aligned}
\tilde{H}_2 &= H + h^2 \left(\frac{1}{12} \left\{\left\{T,V\right\},V\right\} - \frac{1}{24} \left\{ \left\{V,T\right\},T\right\}  \right. \\
&\left. - \frac{1}{4} \sum_i \sum_{j > i} \left( \frac{1}{12} \left\{\left\{T_{i j},V_{i j}\right\},V_{i j}\right\} - \frac{1}{24} \left\{ \left\{V_{i j},T_{i j}\right\},T_{i j}\right\} \right) \right).
\end{aligned}
\end{equation*}
Notice how $\tilde{H}$ has no dependence on the order of pairs in the integrator to second order.  First, we can verify our result that $\tilde{H}$ should only contain terms even in $h$.  By measuring the error in $\tilde{H}_2$, we can verify it scales as $h^4$ as we would predict.  Next, we can verify that $\tilde{H}_2$ should be better conserved than $\tilde{H}$ for small enough $h$; similar experiments were carried out in \cite{engle05}.  To do this, we use the 3BPP again for time $t=1.5$, before the first collision and choose $h = 0.01$.  We indeed find $\tilde{H}_2$ to be conserved better than $H$, as we show in Figure \ref{fig:tildeh}.  

We note the power law increase in global energy error in $\zeta_h^2$.  The initial slope of both curves in Figure \ref{fig:tildeh} is $\approx 2$.  The higher order RK4 shows a slope of $\approx 2.3$ for the same time step.  The slope in Figure \ref{fig:tildeh} is {sensitive to} initial conditions.  In particular, we note that since the 3BPP starts from rest, the Kepler solvers in the first step will all solve orbits with approximately parabolic motion.  If we perturb the velocities in the initial conditions such the orbits are elliptical or hyperbolic, we find the slope of $\zeta_h^2$ to be $\approx 1$ in all cases, i.e. the error is proportional to the number of time steps.  The same holds for the RK4 integrator: the slope is $\approx 1$ for all sets of initial velocities we tried.  

Linear growth in global energy error is typical for non-symplectic integrators as we will also see in Figure \ref{fig:fig8} {below}.  The fact that we see similar error growth in $\zeta_h^2$ is due to the chaotic nature of the problem and is expected \citep{chann90}.  This is not necessarily a cause for concern, and $\phi_h^2$ still yields very positive results for chaotic problems.  We will explain why symplectic integrators can still be well-suited for chaotic problems and study chaotic problems in Section \ref{sec:chaos}.  

Now, we return to the simpler $\phi_h^2$.  {For all tests we tried, $\zeta_h^2$ was less efficient than $\phi_h^2$, so we do not consider it further}.  We can numerically demonstrate the symplecticity of $\phi_h^2$. To do this, we apply  $\phi_h^2$  for one time step to the 3BPP and calculate a Jacobian using Richardson extrapolation \citep{press02}.  We then calculate the right side of eq. \eqref{eq:sympcond} and subtract it from the left side.  This quantity should be 0 to roundoff error for a symplectic method.  We take the absolute value of this difference matrix and sum all the array elements, and call this quantity $dI$.  We compare the result in $dI$ for $\phi_h^2$, KDK leapfrog (we find similar results for DKD leapfrog), a Runge-Kutta 2nd order method (RK2), and { } {SAKURA}.  D stands for drift, defined previously, and K stands for kick.  A kick is defined, as in the case of a drift, for a separable Hamiltonian: $\bm{p}^\prime = \bm{p} + h \bm{f}$, where $\bm{p}^\prime$ is a new momentum, and $\bm{p}$ and $\bm{f}$ are the momentum and force respectively.  We then calculate $dI^{\prime} = dI/m$, with $m$ the number of maps per integrator.  For example, $\phi_h^2$ has $m=14$.       

The results are shown in Table \ref{tab:symplec}.  Even though we calculate $dI$ for only one step, we see a clear difference between the top three symplectic methods and bottom 2 non-symplectic methods.
\begin{center}
\begin{table}
\caption{Symplecticity (measured by $dI^\prime$) comparison of $\phi_h^2$ with other second order methods described in the text after one step in the 3BPP.  The three known symplectic methods show the smallest values of $dI^\prime$.}
\centering
\begin{tabular}{| c || c| c|}
	\hline
	 & $dI^\prime$  \\ [3ex] \hline
	KDK &  $2.0\times 10^{-13}$ \\ \hline
	$\phi_h^2$ &  $2.8\times 10^{-13}$ \\ \hline
	4 KDK steps & $7.9\times 10^{-13}$ \\ \hline
	{SAKURA} & $7.4\times 10^{-11}$ \\ \hline
	RK2 & $1.8\times 10^{-9}$ \\ \hline
	\end{tabular}
\label{tab:symplec}
\end{table}
\end{center}

\section{Numerical tests of method}
\label{sec:numtest}
\subsection{Regular systems}
\label{sec:reg}
We now test the performance of $\phi_h^2$ against standard integrators when applied to $N$-body problems.  {SAKURA is presented in Listing 1 of} \cite{gon14}{ as Python code.  To remain faithful to Listing 1, we implement it and the other codes in Matlab, which has similar syntax to Python.}  We first study regular problems, where the properties of the symplectic integrator are most clearly seen.  The importance of regular problems in testing symplectic integrators is significant.  For small $h$, the KAM tori characteristic of Hamiltonian systems translates into close new invariant KAM tori under symplectic discretization (see \cite{chann90}).  The implications of the near invariant KAM tori can be seen when we study special periodic solutions of the $N$-body problem, as excellent long term behavior for small enough $h$. 

In standard collisional codes, even those with regularization, the lack of a symplectic integrator means there will be secular error growth in the integrals of motion, linear in time for many common methods.  The growth is characteristic of the integrator, and distinct from the growth of roundoff error. 

For our first problem, we study a hirerchical triple problem (HTP).  In this system there is a tight binary initially aligned with the $x$-axis and at apoapsis.  In units where $G=1$, these particles have mass $m_1=m_2=1/2$.  Their semi major axis is $a_0=0.01$ and eccentricity is $e_0=0.9$.  Their orbital period is therefore $P_0 \approx 0.0063$.  Their initial velocities are oriented in the $y$ direction.

There is a third particle with $m_3=1$ that forms a binary pair with the center of mass of the first pair of particles.  This second pair also begins aligned with the $x$ -axis and at apoapsis.  The semi-major axis is $a_1=1$.  The orbit is circular, so the period of this second system is $P_1 \approx 0.14$, which is over $20$ times $P_0$.  The initial velocity of this pair is also in the $y$ direction.  The center of mass position and velocity of the triple system is 0.  The total run time for this system will be $t = 1.4$, which corresponds to about 10 periods of the larger binary and 220 periods of the smaller binary.  During this time, any non-symplectic or non-geometric integrator will show dissipation and secular errors in the integrals of motion.  

\begin{figure*}
 \includegraphics[width=.75\textwidth]{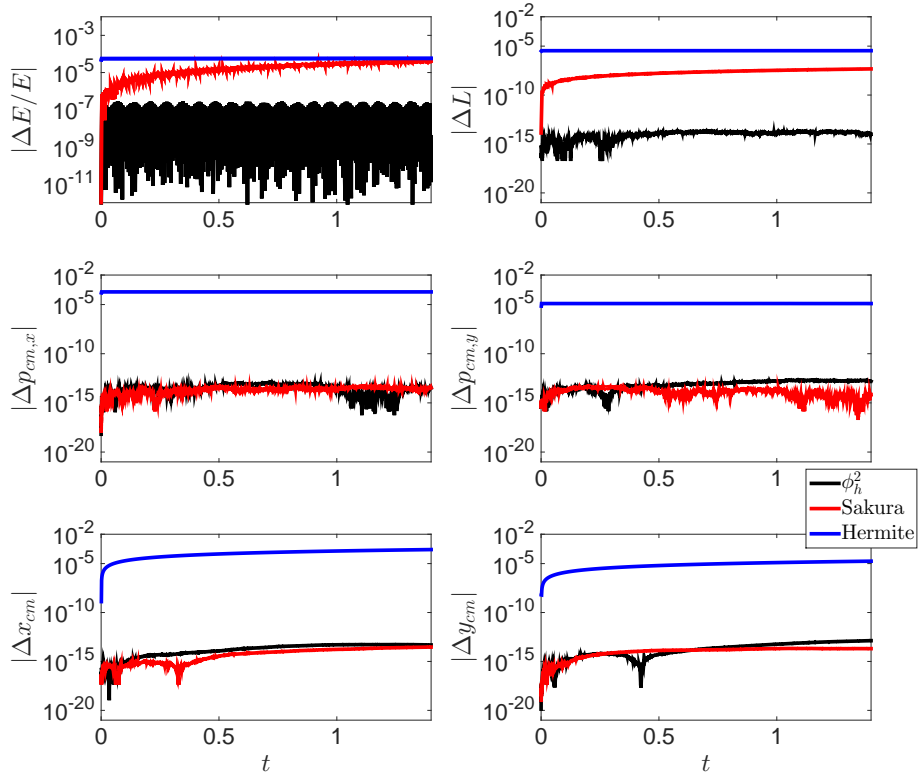}
  \caption{The HTP problem, described in the text, run for $t=1.4$.  The three integrators are normalized to approximately equal cpu effort.  $\phi_h^2$ has the best performance across all integrals of motion and its property that it has a conserved, nearby $\tilde{H}$, is seen by the periodic behavior in energy conservation.}
  \label{fig:hier}
  \centering
\end{figure*}

In Figure \ref{fig:hier} we compare the conserved quantities for $\phi_h^2$, {SAKURA}, and a fourth order Hermite method.  We choose time step parameters such that the cpu time of the integrators is approximately the same ($t_{\mathrm{cpu}} \approx 50$).  Hermite uses $\eta = 10^{-3}$ ($\eta$ is a proxy for the time step, see \cite{aar08}), $\phi_h^2$ uses $h=1.5 \times 10^{-5}$ as does {SAKURA}.  The {f}igure shows the result that no secular growth occurs in any integrals of motion for $\phi_h^2$, and the error in energy is bounded.   $\phi_h^2$ performs better by all proxies compared to the other integrators.

In Figure \ref{fig:hier} we saw the first signs of the bound on energy error due to the combined use of reversible and symplectic methods.   Another periodic solution is the figure-eight three body problem, described mathematically in \cite{chen00}.  This time our goal is to test how important are the properties of symplecticity and time reversibility in long term behavior.  We analyze this problem with the same initial conditions as \cite{chen00} and study its evolution under different integrators.  For this problem, we choose different time steps (and $\eta$ parameter in the Hermite code) for the integrators such that the initial energy error is comparable. 

\begin{figure}
 \includegraphics[width=.5\textwidth]{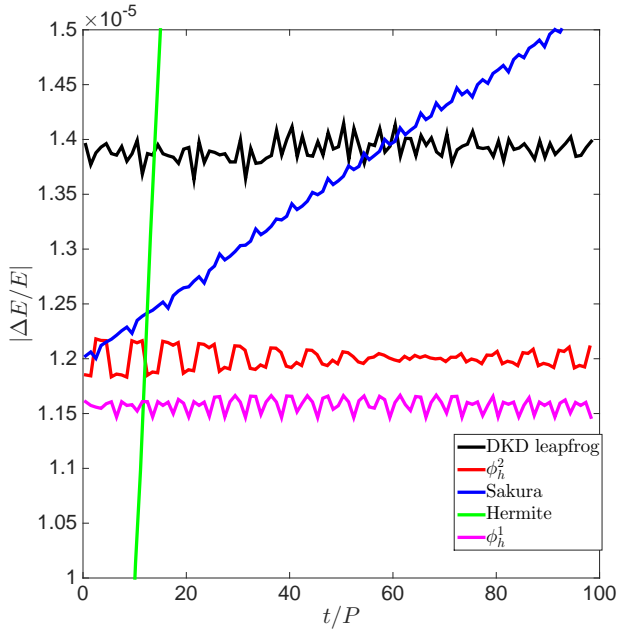}
  \caption{Energy drift for different integrators when applied to the figure-eight three body problem \citep{chen00}.  We choose different time step parameters for the integrators such that the initial energy error is comparable.  There is no drift for the symplectic methods.}
  \label{fig:fig8}
  \centering
\end{figure}

We run the problem for 100 periods.  The period of the problem is approximately $T = 6.32591398$, and in this time there are 7 particle crossings at the origin.  Energy errors are averaged over the period to minimize periodic, expected, variations in energy conservation when plotting.  The result is shown in Figure \ref{fig:fig8}.  
The symplectic methods all have bounded conservation properties over long periods of time as expected.  $\phi_h^1$ is the non-reversible integrator in Algorithm \ref{alg:phih} we used to derive $\phi_h^2$.  Despite $\phi_h^1$ being non-reversible, its good behavior remains.  {SAKURA} shows a linear drift in energy error, as does the Hermite integrator with a larger slope.  They, like RK2 or RK4, have an energy error proportional to the number of time steps taken.    

\subsection{Chaotic systems}
\label{sec:chaos}
Next we show the effect of $\phi_h^2$ when applied to chaotic systems, the more common astrophysical scenario.  Long term conservation of energy is sometimes  not apparent in chaotic systems when a symplectic integrator is used, as we saw in Figure \ref{fig:tildeh}.  This is due to the changing nature of the infinite series in $\tilde{H}$.  Thus, a naive analysis based on energy conservation alone may make one believe that a symplectic integrator performs similarly to standard integrators when chaos is involved.  This analysis would be false.  As explained in \cite{chann90}, symplectic integrators in this case are still highly desirable and preserve structure of the topology of the differential equations by correctly avoiding stable invariant objects.  We can see this result by applying the symplectic Euler map to the simple pendulum problem.  The map in this case is equivalent to the Standard Map (see a discussion and definition of the Standard Map in \cite{yos93}).  While orbits in the Standard Map can be chaotic, for small enough $h$ they avoid invariant topological objects and the chaos is bounded.  

Also relevant are results in \cite{mc93}, where they applied symplectic integrators to chaotic problems.  A particularly interesting result is that they find that the long-term statistics of behavior of orbits converges quite rapidly for the perturbed linear oscillator, a chaotic problem.  This reminds us of a recent result by \cite{pz14} where it is found that statistics of orbits for three-body problems converge to correct answers for accurate enough integrators.  An interesting question is whether the convergence rate improves for a symplectic integrator like $\phi_h^2$, which they did not test. 

We proceed with these results in mind.  Binaries in globular clusters and other collisional astrophysical systems have been the subject of much study and pose one of the greatest challenges to these simulations (\cite{heg03}, \cite{aar08}).  Realistic globular clusters have a fraction of stars in primordial binaries and here we study this effect again in a toy problem.  We choose a problem of 5 bodies sampled from a Plummer sphere.  The properties of the Plummer sphere are described in \cite{heg03}.  Standard units are used in which $G=M=1$ and $E= 1/4$.  $M$ is the total mass.  Each particle has mass $m=1/N$.  This corresponds to a half mass radius $r_h \approx 4/5$.  In these units, the crossing time is $t_{\mathrm{cross}} \approx 1.1$.  

\begin{figure*}
\includegraphics[width=1.1\textwidth]{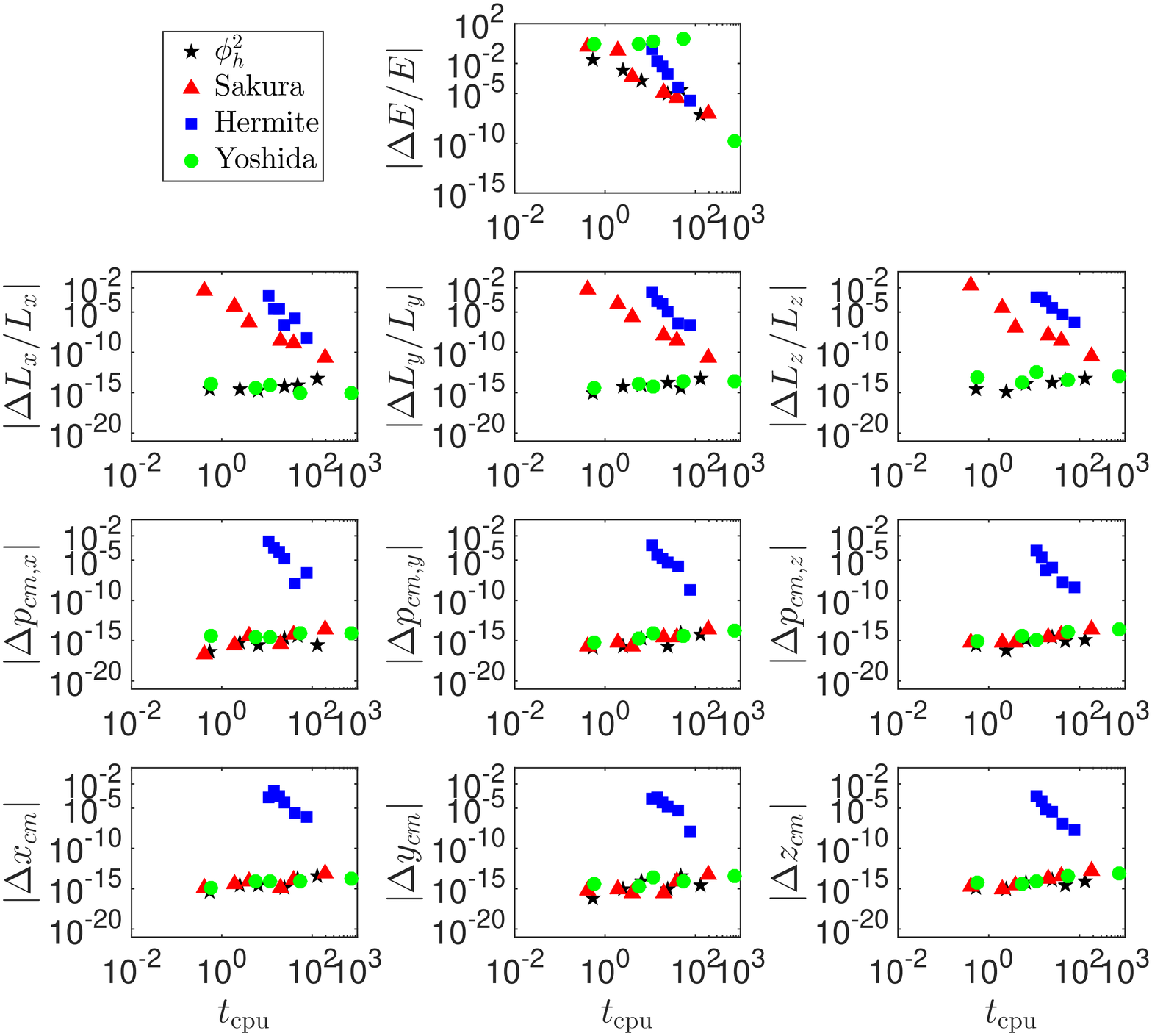}
  \caption{Accuracy achieved for a given computational effort.  5 particles are sampled from a Plummer sphere and a binary is included as described in the text.  The binary is twice as bound as the Plummer sphere in standard units.  {Numerical results are shown for the 10 integrals of motion including 6 center of mass phase space variables}.  $\phi_h^2$ outperforms the other methods.}
  \label{fig:5bodplummerbin}
  \centering
\end{figure*}

Now, we add an extra body that forms a tight binary with one of the original 5 bodies.  The binary has semi major axis $a=0.01$ and eccentricity $e=0.9$.  The mass of each particle in the binary is $1/10$.  In practice this means the binary is twice as bound as the Plummer sphere ($E=-1/2$).  We run this problem for $t/t_{\mathrm{cross}} = 4.5$ with various integrators we have discussed, including the 4th order generalization of leapfrog labelled as `Yoshida' \citep{yos90}.  The result is shown in Figure \ref{fig:5bodplummerbin}.
$\phi_h^2$ outperforms the other methods in conservation properties for a given level of computing effort.  So $\phi_h^2$ is an integrator which shows good performance when primordial binaries are present.  For binaries sufficiently tight and far away from the other particles, $\phi_h^2$ will solve their motion more accurately for a given compute time than the other integrators.  As mentioned in Section \ref{sec:bask}, we can increase the order of our integrator from a second order integrator by the method of \cite{yos90}.  For this problem, tests did not indicate significant accuracy improvements for a given computation time using the fourth order reversible generalization of $\phi_h^2$.  For general Plummer sphere problems without any binary formation we find comparable performance between $\phi_h^2$ and a Hermite method.  However, based on results from \cite{gon14} with {SAKURA}, where they found their method reproduced the cluster radius accurately and more rapidly than other methods for a moderate $N$ Plummer model, we {expect} $\phi_h^2$ {to perform well} for larger Plummer models.  This is especially true since $\phi_h^2$ shows improved performance over {SAKURA} {in all our tests}.

\begin{figure}
 \includegraphics[width=.5\textwidth]{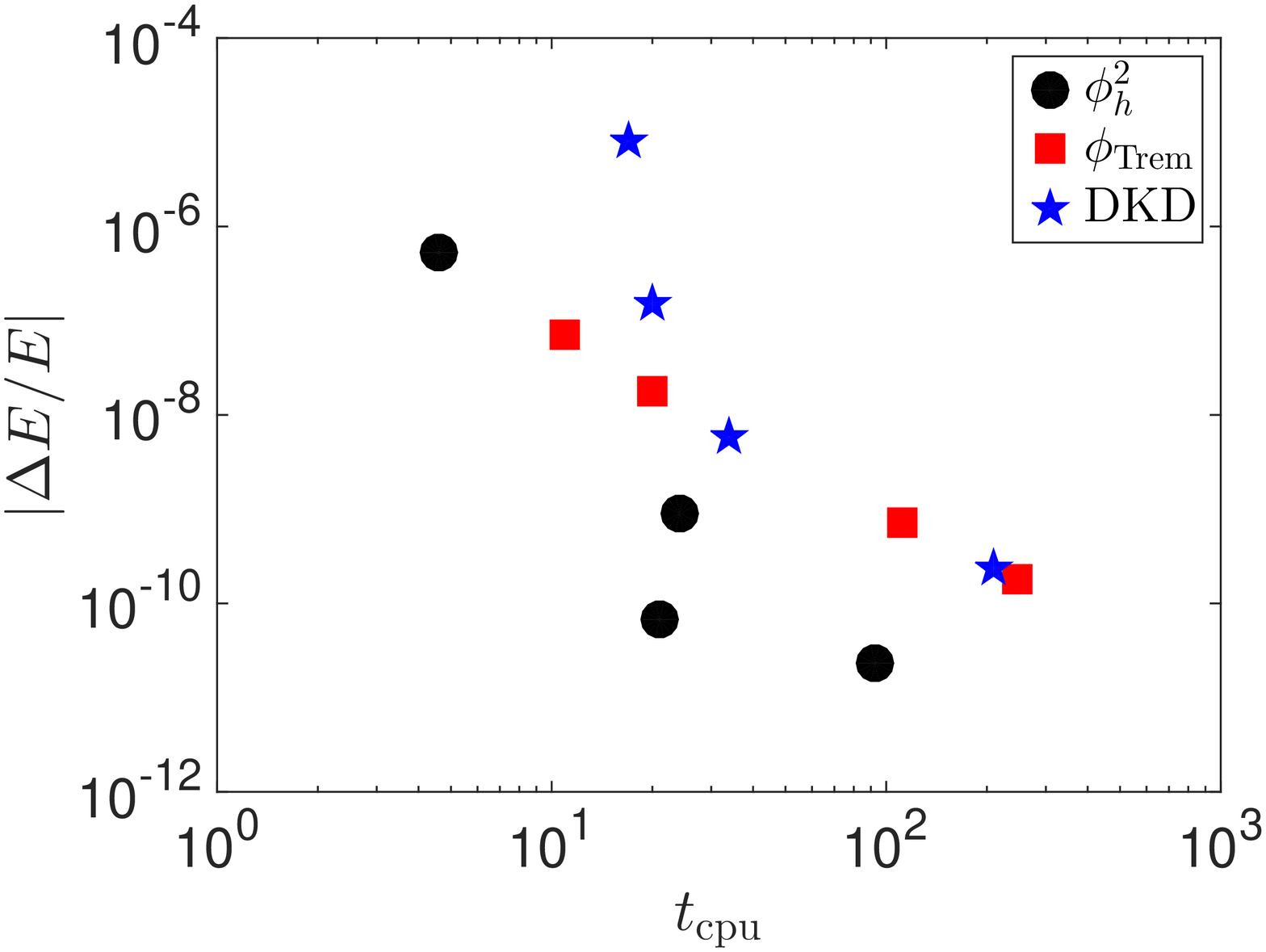}
  \caption{Energy conservation for a given computational effort for the 3BPP.  The problem is run until $t=4$, corresponding to a time shortly after third encounter.  $\phi_{\mathrm{Trem}}$ performs better than leapfrog, but $\phi_h^2$ has the best performance.}
  \label{fig:sympmethods}
  \centering
\end{figure}

We now revisit the 3BPP presented in section \ref{sec:bask}.  The purpose here is to compare the performance of $\phi_h^2$ with the second order symplectic integrator of \cite{mik99} and \cite{preto99}, which we call $\phi_{\mathrm{Trem}}$, and leapfrog.  The time step function for $\phi_{\mathrm{Trem}}$ will be that presented in \cite{mik99}.  {Section} \ref{sec:deriv} {shows leapfrog's conservation of} all integrals of motion, except for the energy, to machine precision.  $\phi_{\mathrm{Trem}}$ also conserves these integrals well, so we will only focus on energy conservation.  We consider the conservation in energy for a given computational effort (the computational effort is changed by varying the timestep).  We run the 3BPP for $t=4$: this corresponds to a time shortly after the third encounter.  {The result is shown in Figure} \ref{fig:sympmethods}.  

We see the poorest integrator is DKD leapfrog, which is not surprising.  Improved performance is shown by $\phi_{\mathrm{Trem}}$, but the best results are obtained by $\phi_h^2$.  The superior performance of $\phi_h^2$ over $\phi_{\mathrm{Trem}}$ was not limited to this problem, but was present in {all Plummer sphere tests with primordial binaries} we tested.  For example, if we take a 5 body Plummer model as described previously in this section, run for $t/t_{\mathrm{cross}} = 2.9$, and calculate {$\delta = \left|\Delta E/E \right|$}, $\phi_h^2$ yields {$t_{\mathrm{cpu}} = 64$} with {$\delta = 9.6 \times 10^{-6}$} while $\phi_{\mathrm{Trem}}$ yields {$t_{\mathrm{cpu}} = 100$} and {$\delta = 1.3 \times 10^{-5}$}  {: better conservation for smaller compute times.}

\section{Symplectic and reversible integrator combining kicks and Kepler solvers}
\label{sec:comb}
{$\phi_h^2$ uses Kepler solutions in place of the faster kick steps of DKD leapfrog.}  We now describe a symplectic and reversible integrator combining the speed of leapfrog and the collisional accuracy of Kepler solvers.  To do this, we write a first order non-reversible symplectic integrator as we did in eq. \eqref{eq:firsto}.  Let $A$ be a set of pairs of particles and $A^C$ is its complement (all the other pairs).  Then the pseudocode for $\phi_h^{\prime}$ is shown in Algorithm \ref{alg:phihp}.
\begin{algorithm} 
\caption{Pseudocode for $\phi_h^{\prime}$}
\label{alg:phihp}
     \SetAlgoLined
     Drift all particles for time $h$ \;
     \For{pairs of particles ($i$, $j$) in $A^C$}{
     		Kick particles $i$ and $j$ using their mutual force only.
     }
     \For{pairs of particles ($i$, $j$) in $A$}{
 		Drift particles $i$ and $j$ for time $-h$ \;
		Apply a Kepler solver to advance the relative coordinates of $i$ and $j$ by $h$ \;
		Advance center of mass coordinates of $i$ and $j$ by $h$ \;
		}
\end{algorithm}
With operators, we have eq. \eqref{eq:firstop}:
\begin{equation}
\begin{aligned}
\phi_h^{\prime} &=  \left( \prod_{(i,j) \epsilon A} \exp \left({h D_{K_{i j}}} \right) \exp \left({- h D_{T_{i j}}} \right) \right)  \exp \left({h D_W} \right) \exp \left({h D_T} \right). \\
&=  \left( \prod_{A} \psi_h^{3ij}  \psi_{-h}^{2ij} \right)\psi_h^4 \psi_h^1,
\end{aligned}
\label{eq:firstop}
\end{equation}
with
\begin{equation*}
W = \sum_{(i,j) \epsilon A^C} V_{i j}.
\end{equation*}
With method $\phi_h^{\prime}$ we could choose, for example, to kick pairs that are far from each other and apply the Kepler solver to tight binaries.  In practice tight binaries will interact with other particles and perhaps break up, and this integrator may not be practical for such situations.  A second order method is developed following the step of eq. \eqref{eq:phi2}:
\begin{equation}
\phi_h^{2 \prime} = \phi_{h/2}^{\prime \dag} \phi_{h/2}^{\prime}.
\label{eq:phi2p}
\end{equation}
{As we showed in Section} \ref{sec:deriv} {, a kick step is symplectic and conserves 9 of 10 integrals of motion to machine precision.}

\begin{figure}
\includegraphics[width=.5\textwidth]{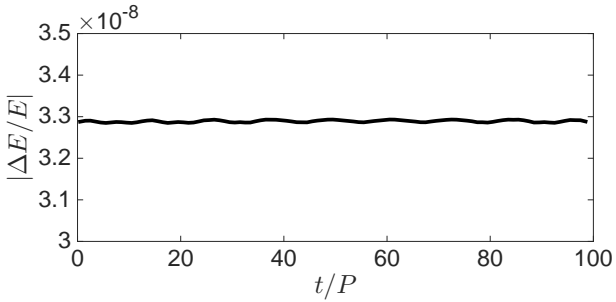}
  \caption{$\phi_h^{2 \prime}$ applied to the figure-eight three body problem \citep{chen00}.  A time step $h=0.001$ is used and the problem is run for 100 periods.  The energy error is averaged over the period as in Figure \ref{fig:fig8}.  Consistent with existence of a nearby $\tilde{H}$, and periodic behavior characteristic of a symplectic integrator, there is no secular drift in the energy error.}
  \label{fig:fig8p}
  \centering
\end{figure}

We first show the long-term conservation properties characteristic of symplectic and reversible methods of $\phi_h^{2\prime}$.  We apply the integrator to the figure-eight three body problem.  We assign the three particles to numbers 1-3.  We let $A = \{ (1,2), (1,3) \}$, where $(1,2)$ and $(1,3)$ are the pairs of particles evolved with Kepler solvers, and $A^C = \{ (2,3) \}$.  We choose $h=0.001$ and run the problem for 100 periods.  The result is shown in Figure \ref{fig:fig8p} and we indeed see the excellent conservation properties we expect from symplectic and reversible methods.  We have again averaged the energy error over each period.  

Now we apply the integrator to a more complicated problem.  We take the same figure-eight problem, but now replace each particle by a binary whose center of mass behaves identically as in the original case.  The binary pairs will have $e=0.9$ and $a=0.01$ again as in the case of Figure \ref{fig:5bodplummerbin}.  The binaries have total energy 29 times greater in magnitude than the energy of the original problem.  They complete more than 1000 orbits each per each period of the original problem.  In this case it is not clear whether the KAM theorem will apply, but numerical experiments indicate that for a modest number of periods, the system remains close to periodic.

\begin{figure*}
 \includegraphics[width=.75\textwidth]{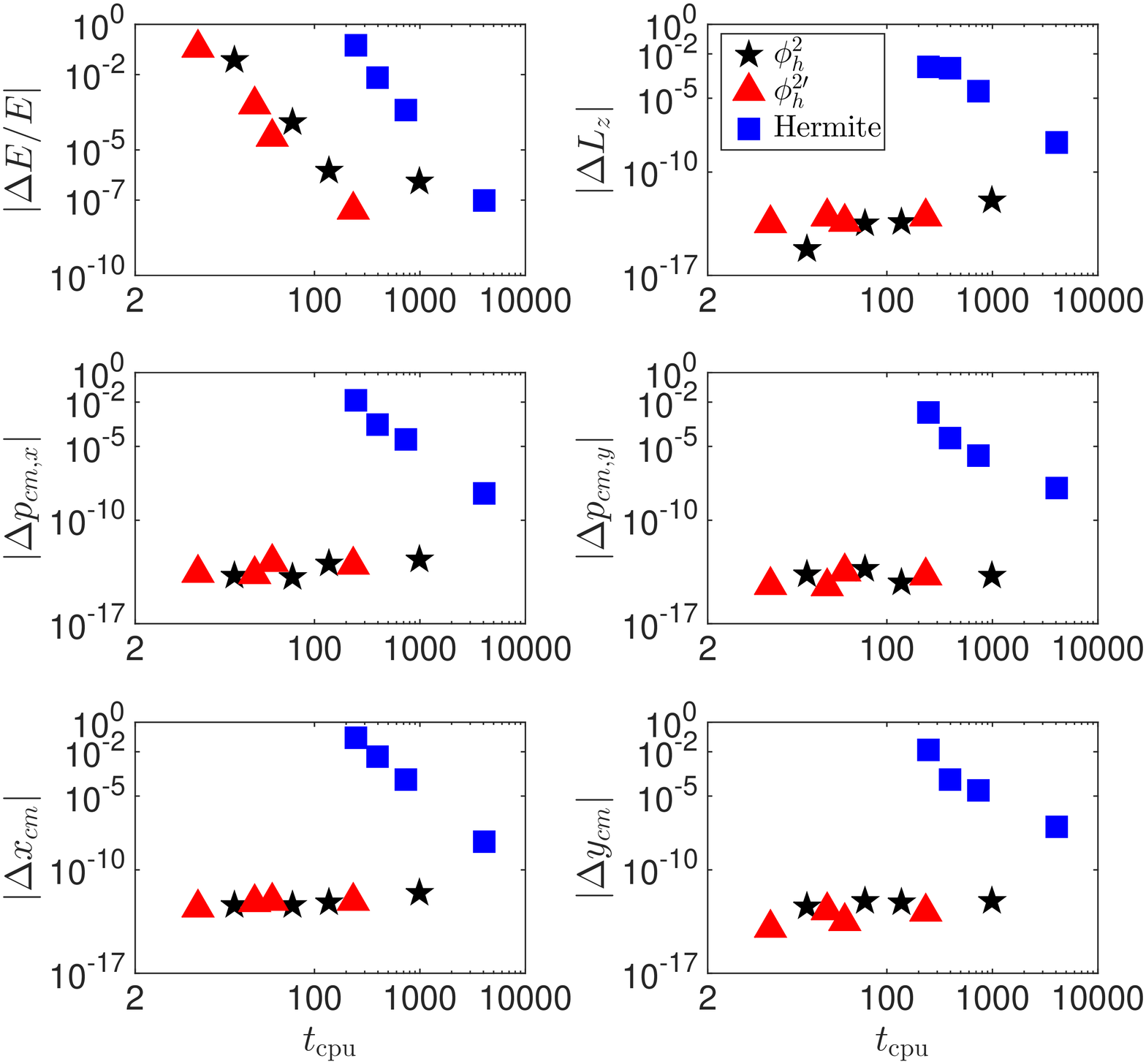}
  \caption{Level of computing effort required for a given accuracy for different integrators applied to the figure-eight three body problem \citep{chen00} but with each particle now replaced with a tight binary, as described in the text.  The problem is run for 2 periods of the original three body problem. $\phi_h^{2\prime}$ {is the best method for energy conservation and performs at machine precision in conservation of the other integrals.}}
  \label{fig:phi2hp}
  \centering
\end{figure*}

We measure the computational effort required to achieve a given accuracy in the integrals of motion.  The system is run for 2 periods.  The result is shown in Figure \ref{fig:phi2hp}.  In this case $\phi_h^{2\prime}$ performs {slightly} better than $\phi_h^2$ {in energy conservation,} indicating it {is also} a good collisional integration tool.  {Conservation of other integrals is at machine precision, as is the case of $\phi_h^2$; we can show this through similar analysis to that of map K in section} \ref{sec:deriv}. 
\section{Conclusion}
\label{sec:conc}
The {main} goal of this paper is to present {a new} collisional $N$-body symplectic method and its various properties and perform tests on toy problems.  In the small problems we present, it performs equally as well or better than the standard Hermite method and the symplectic method of \cite{preto99} and \cite{mik99}.  We discuss, prove, and numerically show many of the integrator's properties: its symplecticity, its exact conservation of 9 out of 10 integrals of motion, its reversibility, its order of integration, and its suitability for collisional problems.  We perform tests on regular three body problems and chaotic $N$-body problems.  Although we have only presented a second order integrator, the order can be increased by the method of \cite{yos90}.  

{A secondary goal is to} present a method such that not every pair needs to be treated via Kepler solver, and as a result we can increase speed for a given accuracy in some problems.

The next step, which we will take in forthcoming work, will be to carry out larger $N$ tests on more realistic problems.  We are especially optimistic about the tests because of results by \cite{gon14}, {since all our tests show similar or better performance than SAKURA}.  The integrator should prove a useful tool in the study of collisional $N$-body dynamics.   

\section{Acknowledgements}
\label{sec:ack}
{It is a pleasure to thank S.F. Portegies Zwart and T. Boekholt for numerous discussions regarding SAKURA.  We thank Katherine Deck for providing a Kepler solver.  We thank Brendan Griffen for useful discussions.}  DMH acknowledges support by the National Science Foundation Graduate Research Fellowship under Grant No. 1122374. 

\bibliographystyle{mn2e}
\bibliography{doc}

\onecolumn
\appendix
\section{{Time reversibility of SAKURA}}
\label{app}

Here we show that, for $N>2$, the SAKURA algorithm presented in {Listing 1} of \cite{gon14} is not time reversible.  We then provide an explanation for the irreversibility.

Initial conditions at time $t$ are $\bm{y} = \{{\bm r}_i,{\bm v}_i\}$.  $\bm{r}_i$ and $\bm{v}_i$ are position and velocity vectors, respectively, for particle $i$.  One timestep transforms these initial conditions into final conditions at time $t+h$:
\begin{eqnarray}\label{Sakura}
  {\bm r}'_i&=&{\bm r}_i+h{\bm v}_i+\sum_{k\ne i}\frac{m_k}{m_i+m_k}\left[{\bm K}({\bm r}_i-
    {\bm r}_k,{\bm v}_i-{\bm v}_k,m_i+m_k,h)-({\bm r}_i+h{\bm v}_i)+({\bm r}_k+h{\bm v}_k)
    \right]\ ,\nonumber\\
  {\bm v}'_i&=&{\bm v}_i+\sum_{k\ne i}\frac{m_k}{m_i+m_k}\left[\dot{\bm K}({\bm r}_i-{\bm r}_k,
    {\bm v}_i-{\bm v}_k,m_i+m_k,h)-{\bm v}_i+{\bm v}_k\right]\ .
\end{eqnarray}
Equation \eqref{Sakura} corresponds exactly to {Listing 1}.  Here, ${\bm K}({\bm r},{\bm v},m,t)={\bm x}(t)$ is the solution to the Kepler problem
\begin{equation}
  \frac{d^2{\bm x}}{dt^2}=-\frac{Gm{\bm x}}{|{\bm x}|^3}\nonumber
\end{equation}
subject to initial conditions ${\bm x}={\bm r}, $ $d{\bm x}/dt={\bm v}$ at $t=0$. The Kepler solution itself is time reversible; the question is whether algorithm (\ref{Sakura}) is.

To determine time reversibility, form the pair differences:
\begin{eqnarray}\label{pair}
  {\bm r}'_i-{\bm r}'_j&=&{\bm K}({\bm r}_i-{\bm r}_j,{\bm v}_i-{\bm v}_j,m_i+m_j,h)\nonumber\\
    &&+\sum_{k\ne i,j}\frac{m_k}{m_i+m_k}\left[{\bm K}({\bm r}_i-{\bm r}_k,{\bm v}_i-{\bm v}_k,
      m_i+m_k,h)-({\bm r}_i+h{\bm v}_i)+({\bm r}_k+h{\bm v}_k)\right]\nonumber\\
    &&-\sum_{k\ne i,j}\frac{m_k}{m_j+m_k}\left[{\bm K}({\bm r}_j-{\bm r}_k,{\bm v}_j-{\bm v}_k,
      m_j+m_k,h)-({\bm r}_j+h{\bm v}_j)+({\bm r}_k+h{\bm v}_k)\right]\ ,\nonumber\\
    {\bm v}'_i-{\bm v}'_j&=&\dot{\bm K}({\bm r}_i-{\bm r}_j,{\bm v}_i-{\bm v}_j,m_i+m_j,h)
    \nonumber\\
    &&+\sum_{k\ne i,j}\frac{m_k}{m_i+m_k}\left[\dot{\bm K}({\bm r}_i-{\bm r}_k,
      {\bm v}_i-{\bm v}_k,m_i+m_k,h)-{\bm v}_i+{\bm v}_k\right]\nonumber\\
    &&-\sum_{k\ne i,j}\frac{m_k}{m_j+m_k}\left[\dot{\bm K}({\bm r}_j-{\bm r}_k,
      {\bm v}_j-{\bm v}_k,m_j+m_k,h)-{\bm v}_j+{\bm v}_k\right]\ .
\end{eqnarray}
These difference vectors are redundant coordinates and do not form a complete coordinate system.  We can complete the coordinate system using the center of mass position and the total momentum.  We can show using Equations \eqref{Sakura} that SAKURA conserves exactly the center of mass integrals.  This result is supported numerically by Figures \ref{fig:hier} and \ref{fig:5bodplummerbin}.  It follows SAKURA's center of mass motion is reversible.  Consider the case $N=2$, for which
\begin{eqnarray}\label{n=2}
  {\bm r}'_1-{\bm r}'_2&=&{\bm K}({\bm r}_1-{\bm r}_2,{\bm v}_1-{\bm v}_2,m_1+m_2,h)\ ,
    \nonumber\\
  {\bm v}'_1-{\bm v}'_2&=&\dot {\bm K}({\bm r}_1-{\bm r}_2,{\bm v}_1-{\bm v}_2,m_1+m_2,h)\ .
\end{eqnarray}
Clearly, this is time reversible,
\begin{eqnarray}\label{reverse2}
  {\bm r}_1-{\bm r}_2&=&{\bm K}({\bm r}'_1-{\bm r}'_2,{\bm v}'_1-{\bm v}'_2,m_1+m_2,-h)\ ,
    \nonumber\\
  {\bm v}_1-{\bm v}_2&=&\dot {\bm K}({\bm r}'_1-{\bm r}'_2,{\bm v}'_1-{\bm v}'_2,m_1+m_2,-h)\ .
\end{eqnarray}
Thus the total solution is reversible.  It is also symplectic because it solves the two-body problem exactly.

However, $N>2$ is not time reversible, for example, $N=3$:
\begin{eqnarray}\label{n=3}
  {\bm r}'_1-{\bm r}'_2&=&{\bm K}({\bm r}_1-{\bm r}_2,{\bm v}_1-{\bm v}_2,m_1+m_2,h)
    \nonumber\\
    &&+\frac{m_3}{m_1+m_3}\left[{\bm K}({\bm r}_1-{\bm r}_3,{\bm v}_1-{\bm v}_3,
      m_1+m_3,h)-({\bm r}_1+h{\bm v}_1)+({\bm r}_3+h{\bm v}_3)\right]\nonumber\\
    &&-\frac{m_3}{m_2+m_3}\left[{\bm K}({\bm r}_2-{\bm r}_3,{\bm v}_2-{\bm v}_3,
      m_2+m_3,h)-({\bm r}_2+h{\bm v}_2)+({\bm r}_3+h{\bm v}_3)\right]\ ,\nonumber\\
  {\bm v}'_1-{\bm v}'_2&=&\dot {\bm K}({\bm r}_1-{\bm r}_2,{\bm v}_1-{\bm v}_2,m_1+m_2,h)
    \nonumber\\
    &&+\frac{m_3}{m_1+m_3}\left[\dot{\bm K}({\bm r}_1-{\bm r}_3,{\bm v}_1-{\bm v}_3,
      m_1+m_3,h)-{\bm v}_1+{\bm v}_3\right]\nonumber\\
    &&-\frac{m_3}{m_2+m_3}\left[\dot{\bm K}({\bm r}_2-{\bm r}_3,{\bm v}_2-{\bm v}_3,
      m_2+m_3,h)-{\bm v}_2+{\bm v}_3\right]\ .
\end{eqnarray}
While this is more complicated than (\ref{n=2}), it is still possible that the algorithm is time reversible; to be sure, we must evaluate the other two pairs and then explicitly test time reversal.
\begin{eqnarray}\label{n=3a}
  {\bm r}'_1-{\bm r}'_3&=&{\bm K}({\bm r}_1-{\bm r}_3,{\bm v}_1-{\bm v}_3,m_1+m_3,h)
    \nonumber\\
    &&+\frac{m_2}{m_1+m_2}\left[{\bm K}({\bm r}_1-{\bm r}_2,{\bm v}_1-{\bm v}_2,
      m_1+m_2,h)-({\bm r}_1+h{\bm v}_1)+({\bm r}_2+h{\bm v}_2)\right]\nonumber\\
    &&+\frac{m_2}{m_2+m_3}\left[{\bm K}({\bm r}_2-{\bm r}_3,{\bm v}_2-{\bm v}_3,
      m_2+m_3,h)-({\bm r}_2+h{\bm v}_2)+({\bm r}_3+h{\bm v}_3)\right]\ ,\nonumber\\
  {\bm v}'_1-{\bm v}'_3&=&\dot {\bm K}({\bm r}_1-{\bm r}_3,{\bm v}_1-{\bm v}_3,m_1+m_3,h)
    \nonumber\\
    &&+\frac{m_2}{m_1+m_2}\left[\dot{\bm K}({\bm r}_1-{\bm r}_2,{\bm v}_1-{\bm v}_2,
      m_1+m_2,h)-{\bm v}_1+{\bm v}_2\right]\nonumber\\
    &&+\frac{m_2}{m_2+m_3}\left[\dot{\bm K}({\bm r}_2-{\bm r}_3,{\bm v}_2-{\bm v}_3,
      m_2+m_3,h)-{\bm v}_2+{\bm v}_3\right]\ ,\nonumber\\
  {\bm r}'_2-{\bm r}'_3&=&{\bm K}({\bm r}_2-{\bm r}_3,{\bm v}_2-{\bm v}_3,m_2+m_3,h)
    \nonumber\\
    &&-\frac{m_1}{m_1+m_2}\left[{\bm K}({\bm r}_1-{\bm r}_2,{\bm v}_1-{\bm v}_2,
      m_1+m_2,h)-({\bm r}_1+h{\bm v}_1)+({\bm r}_2+h{\bm v}_2)\right]\nonumber\\
    &&+\frac{m_1}{m_1+m_3}\left[{\bm K}({\bm r}_1-{\bm r}_3,{\bm v}_1-{\bm v}_3,
      m_1+m_3,h)-({\bm r}_1+h{\bm v}_1)+({\bm r}_3+h{\bm v}_3)\right]\ ,\nonumber\\
  {\bm v}'_2-{\bm v}'_3&=&\dot {\bm K}({\bm r}_2-{\bm r}_3,{\bm v}_2-{\bm v}_3,m_2+m_3,h)
    \nonumber\\
    &&-\frac{m_1}{m_1+m_2}\left[\dot{\bm K}({\bm r}_1-{\bm r}_2,{\bm v}_1-{\bm v}_2,
      m_1+m_2,h)-{\bm v}_1+{\bm v}_2\right]\nonumber\\
    &&+\frac{m_1}{m_1+m_3}\left[\dot{\bm K}({\bm r}_1-{\bm r}_3,{\bm v}_1-{\bm v}_3,
      m_1+m_3,h)-{\bm v}_1+{\bm v}_3\right]\ .
\end{eqnarray}
The presence of the additional pair terms on the right-hand sides of equations (\ref{n=3}) and (\ref{n=3a}) breaks time reversibility.

The reason why SAKURA is not time reversible is that the Kepler steps in (\ref{Sakura}) are not done in serial but instead are done in parallel, using the initial phase space coordinates for every Kepler evaluation. When $N>2$, so that more than one pair needs to be advanced, the phase space variables need to be updated after each Kepler evaluation. This is a necessary but not sufficient condition for time reversibility. An additional requirement is that the pairs be evaluated in a reversible order, which is accomplished in our paper by splitting the timestep in half and using one pair order for the first half, and the reversed pair order for the second half (this is what is meant by the adjoint map).

\end{document}